\documentclass[preprint]{emulateapj}
\usepackage{url}

\slugcomment{ }
\usepackage[breaklinks,colorlinks,urlcolor=blue,citecolor=blue,linkcolor=blue]{hyperref}

\shorttitle{Dense core properties in G14.2 }
\shortauthors{Ohashi et al.}

\begin{document}


\title{Dense core properties in the Infrared Dark cloud G14.225-0.506 revealed by ALMA}


\author{Satoshi Ohashi\altaffilmark{1,2},
Patricio Sanhueza\altaffilmark{2},
Huei-Ru Vivien Chen\altaffilmark{3,4},
Qizhou Zhang\altaffilmark{5},
Gemma Busquet\altaffilmark{6},
Fumitaka Nakamura\altaffilmark{2},
Aina Palau\altaffilmark{7},
and
Ken'ichi Tatematsu\altaffilmark{2,8}
}

\altaffiltext{1}{Department of Astronomy, The University of Tokyo, Bunkyo-ku, Tokyo 113-0033, Japan}
\email{satoshi.ohashi@nao.ac.jp}
\altaffiltext{2}{National Astronomical Observatory of Japan, National Institutes of Natural Sciences, 2-21-1 Osawa, Mitaka, Tokyo 181-8588, Japan}
\altaffiltext{3}{Institute of Astronomy and Department of Physics, National Tsing Hua University, 101, Sec. 2, Kuang Fu Road, Hsinchu 30013, Taiwan} %
\altaffiltext{4}{Academia Sinica Institute of Astronomy and Astrophysics, P.O. Box 23-141, Taipei, 10617, Taiwan} %
\altaffiltext{5}{Harvard-Smithsonian Center for Astrophysics, 60 Garden Street, Cambridge, MA 02318, USA} %
\altaffiltext{6}{Institut de Ci\`encies de l'Espai (IEEC-CSIC), Campus UAB, Carrer de Can Magrans, S/N. E-08193, Barcelona, Catalunya, Spain}
\altaffiltext{7}{Instituto de Radioastronom\'ia y Astrof\'isica, Universidad Nacional Aut\'onoma de M\'exico, P.O. Box 3-72, 58090 Morelia, Michoac\'an, M\'exico}
\altaffiltext{8}{Department of Astronomical Science, SOKENDAI (The Graduate University for Advanced Studies), 2-21-1 Osawa, Mitaka, Tokyo 181-8588, Japan}


\begin{abstract}

We have performed a dense core survey toward the Infrared Dark Cloud G14.225-0.506 at 3 mm continuum emission with the Atacama Large Millimeter/Submillimeter Array (ALMA). This survey covers the two hub-filament systems with an angular resolution of $\sim3\arcsec$ ($\sim0.03$ pc). We identified 48 dense cores. Twenty out of the 48 cores are protostellar due to their association with young stellar objects (YSOs)  and/or X-ray point-sources, while the other 28 cores are likely prestellar and unrelated with known  IR or X-ray emission. Using APEX 870 $\mu$m continuum emission, we also identified the 18 clumps hosting these cores. Through virial analysis using the ALMA N$_2$H$^+$ and VLA/Effelsberg NH$_3$ molecular line data, we found a decreasing trend in the virial parameter with decreasing scales from filaments to clumps, and then to cores. The virial parameters of $0.1-1.3$ in cores, indicate that cores are likely undergoing dynamical collapse. 
  The cumulative Core Mass Function (CMF) for the prestellar cores candidates has a power law index of $\alpha=1.6$, with masses ranging from 1.5 to 22 $M_\odot$. We find no massive prestellar or protostellar cores. Previous studies suggest that massive O-tpye stars have not been produced yet in this region. Therefore, high-mass stars should be formed in the prestellar cores by accreting a significant amount of gas from the surrounding medium. Another possibility is that low-mass  YSOs become massive by accreting from their parent cores that are fed by filaments. These two possibilities might be consistent with the scenario of global hierarchical collapse.

\end{abstract}





\section{Introduction}
The majority of stars are thought to be produced through the formation of clusters \citep{lad03}. In particular, high-mass stars, which have significant effects on the interstellar medium by their strong UV radiation and supernovae, are formed in clusters. High-mass star formation is one of the most outstanding issues in astronomy. The formation process of these stars is still obscure compared with low-mass star formation. For example, it is still unclear whether high-mass stars are formed earlier or later relative to low-mass stars. High-mass stars form in crowded environments (clusters) at large distances ($\gtrsim2$ kpc), making difficult to resolve individual objects in clusters.   Another difficulty is their very short prestellar phase \citep{mot07} in spite of the fact that revealing the prestellar phase is very important to investigate high-mass star-forming processes.

\begin{figure*}[htbp]
  \begin{center}
  \includegraphics[width=16cm]{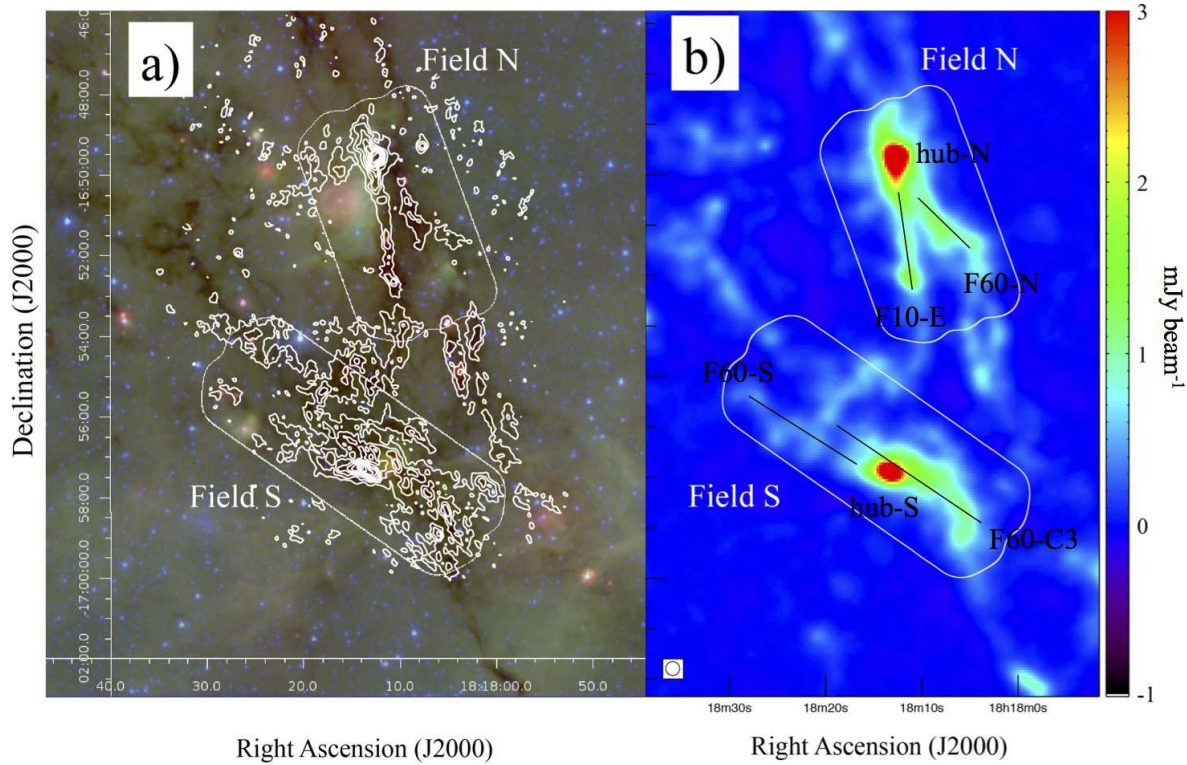}
  \end{center}
  \caption{(Left) Three color images taken by {\it Spitzer} (red/green/blue = 24/8/3.6 $\mu$m) show the G14.2 region. The white contours show the NH$_3$ ($J=1,1$) velocity-integrated intensities from \citet{bus13}. The
lowest contour and the contour step are 3$\sigma$. The 1$\sigma$ noise level for the contour is 0.024 Jy km s$^{-1}$. The integrated velocity range for the molecular line is from 20 to 28 km s$^{-1}$. The white  boundaries show the observed field of ALMA.  (Right) APEX-LABOCA 870 $\mu$m continuum emission, indicating the most prominent filaments with black lines (Busquet et al. 2013). The APEX beam of 22$\arcsec$ is shown in the bottom-left corner of the image.}
  \label{map}
\end{figure*}

Interestingly, filamentary structures are often discovered in Infrared Dark Clouds (IRDCs) and many star-forming regions \citep[e.g.,][]{rat06,san10,and10,arz11,hil11,san13,con16}. 
Filamentary molecular cloud formation has been simulated by colliding flows in the Warm Neutral Medium \citep{vaz07,gom14}.
Recent magneto-hydrodynamical simulations suggest that molecular clouds are formed by accretion of HI clouds through thermal instability and  their structures are filamentary with axes perpendicular to the mean magnetic field lines \citep{ino12,inu15}.

Young stellar object (YSO) groups are often associated with dense clumps at the convergence of multiple filaments, namely ``hub-filament'' systems \citep{mye09,hen12,liu12}.
The morphology of hub-filament systems can be related to the initial strength and configuration of the magnetic field \citep{van14}.    
These young embedded clusters are particularly useful for investigating the origin of the stellar initial mass function (IMF).  Due to their youth, most protostars have not yet reached their final stellar mass.  The mass distribution of protocluster is often referred to as YSO mass function (YMF), which may be altered by subsequent accretion from their parent cores.
Actually, different  YMF and IMF shapes have been reported in IRDCs \citep{pov10,pov16}.

IMF studies began with \citet{sal55}, who measured the IMF from field stars in the solar vicinity and found a power-law form  given by $dN/dM\propto M^{-\gamma}$ with $\gamma=2.35$ for $1-10$ $M_\odot$. Recent observations with high-sensitivity instruments in infrared allow us to identify embedded young clusters and precisely derive the IMF. \citet{mue02} derived the Orion Trapezium cluster IMF from the $K$-band luminosity function and found $\gamma=2.1$ above 0.6 $M_\odot$.

On studies of the mass function of IRDCs, \citet{rag09} have identified clumps ($\sim0.1$ pc) and investigated their mass distribution. They found a slope of 1.76 for masses from 30 to 3000 $M_\odot$. \citet{rat06} found a slope of 2.1 above $\sim100$ $M_\odot$. These power-law indices are consistent with that of the IMF.   It has also been suggested that the CMF is similar in shape to the IMF \citep{nak95,mot98,mot01,alv07,kon10,rad12}. Similar shapes of mass function for IRDC clumps/cores and stars may suggest that clumps and cores lead to the origin of the IMF. However, if these clumps and cores fragment further, the mass spectrum may steepen and change shape at small core scales.  In recent ALMA observations, \citet{zha15} investigated the mass distributions of dense cores in the massive filamentary IRDC G28.34$+$0.06, clump P1, with subsolar mass sensitivities. However, they found a lack of a low-mass population compared with the Salpeter IMF. 
Whether the CMF is related with the YMF or  IMF remains an open question.  More observational data will help to develop a theory for cluster formation.  
The relation between the CMF and IMF can allow us to improve our understanding of star formation. \citet{alv07} have shown a CMF with a single power law and similar to the IMF, but scaled to a higher mass by a factor of about 3. They suggest that the IMF is the direct product of the CMF, which means stars are formed in dense cores one by one at a uniform star-formation efficiency of 30\%. 
It is quite important to investigate the relation of the CMF, YMF, and IMF in high-mass star-forming regions such as hub-filament systems.
  
The IRDC G14.225-0.506 (hereafter G14.2), also known as M17 SWex \citep{pov10,pov16}, is one of the prominent IRDCs in the Galaxy, which is  located southwest of the M17 region  (Figure \ref{map}).
M17 is a well-known HII region excited by the high-mass cluster NGC 6618 \citep{chi08}. 
\citet{elm76} discovered these molecular clouds as an extended very large massive molecular cloud. 
Recent measurements of parallaxes and proper motions using CH$_3$OH masers determine a distance of 1.98 kpc \citep{xu11,wu14}. 
 \citet{bus13} observed this region in NH$_3$ (1,1) and (2,2) lines with the Very Large Array (VLA) and Effelsberg 100m telescope. They identified a network of filaments constituting two hub-filament systems (Figure \ref{map}). The average rotation temperature in these hub regions was derived to be $\sim$ 15 K. 
\citet{pov10} performed a YSO survey toward this region by analyzing the {\it Spitzer} GLIMPSE and MIPSGAL data.  They found that the YMF is steeper than the universal IMF, indicating a deficit of high-mass YSOs.  Subsequently, \citet{pov16} improved the YSO catalog by adding {\it Chandra X-ray} and UKIDSS Galactic Plane Survey observations. As a result, they found a rich population of YSOs and X-ray-emittting intermediate-mass pre-main-sequence stars.
In spite of the large number of intermediate-mass YSOs detected by \citet{pov10} and \citet{pov16}, massive O-type stars are still absent in G14.2.

In this paper, we report ALMA observations of the 3 mm continuum emission. The ALMA data also include molecular lines, which will be presented in a forthcoming paper (Chen et al. 2016, in prep).

\section{Observations}
G14.2 was observed with the ALMA 12-m array on 2015 April 25 in the C34-2/1 configuration with a total of 37 antennas and with the ACA (7-m Array antennas) on 2015 April $29-30$ and May 4 with a total of 10 antennas (Cycle 2 program, Project ID:2013.1.00312.S; PI: Vivien Chen). The observed area is shown in Figure \ref{map}. The total number of 12-m pointings is 58 in field N and 68 in field S.  Including time for calibration, the total observing time was about 1.7 hr for the 12-m Array.  The observations employed the Band 3 (103 GHz for continuum) receivers and the system temperatures ranged from 60 to 90 K.
The projected baselines range from 10 to 346 m (including 7-m and 12-m arrays). Following equation A5 of \citet{pal10}, the largest angular scale (LAS)  recoverable by the 12 and 7-m arrays is 26$\arcsec$ ($\sim0.25$ pc).
The quasar J1733-1304 and J1924-2914 were observed for bandpass, phase, and amplitude calibration. Flux calibration was performed using Neptune and Ceres. The uncertainty of absolute flux calibration is 5\% in Band 3 according to ALMA Cycle 2 Technical Handbook.

The reduction and calibration of the data were done with CASA version 4.3.1 \citep{mcm07} in standard manners, and the data were delivered from the East Asia ALMA Regional Center. ALMA 12-m Array and ACA data sets were combined in the uv plane in CASA. All images of continuum emission were reconstructed with the CASA task CLEAN using natural weighting. The achieved synthesized beams were $3\farcs7\times2\farcs0$ in field N and  $3\farcs1\times2\farcs1$ in field S. The rms noise level for the combined data is 0.2 mJy beam$^{-1}$ before the primary beam correction. The pixel size was set to $0\farcs4$.





\section{Results}
\begin{figure*}[htbp]
  \begin{center}
  \includegraphics[width=16cm]{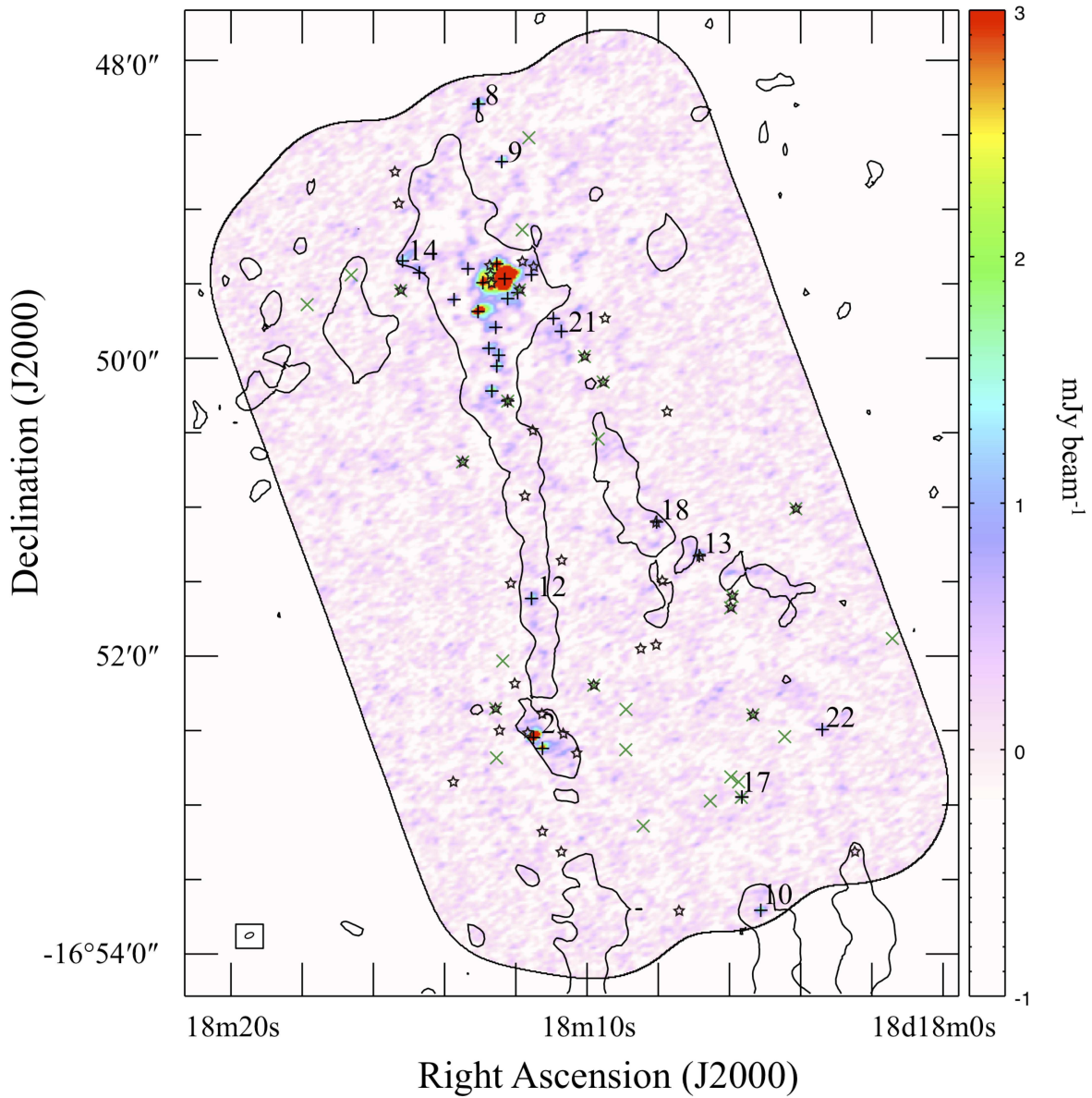}
  \end{center}
  \caption{Field N: ALMA 3 mm continuum image of G14.2. The black crosses indicate all positions of the ALMA cores and the numbers show some positions of the ALMA cores listed in table \ref{alma}.   The stars and the green crosses indicate the locations of the YSOs and the X-ray sources with $A_V > 20$ mag identified by \citet{pov16}, respectively. The black contour shows 4$\sigma$ of the NH$_3$ ($J=1,1$) velocity-integrated intensity. The 1$\sigma$ noise level for the contour is 0.024 Jy km $^{-1}$. The ALMA beam of $3\farcs7\times2\farcs0$  is shown in the bottom-left corner.}
  \label{fig1}
\end{figure*}

\begin{figure*}[htbp]
  \begin{center}
  \includegraphics[width=16cm]{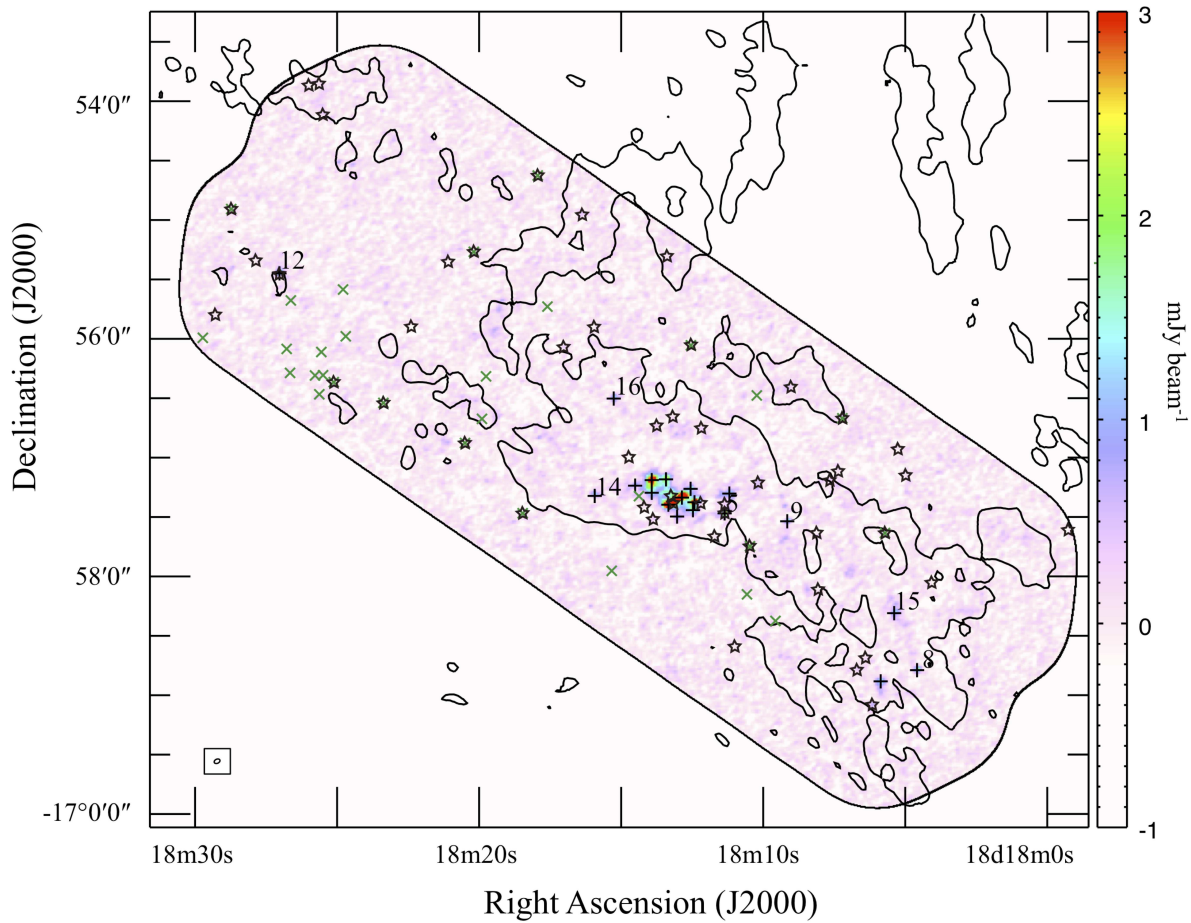}
  \end{center}
  \caption{Same as Figure \ref{fig1} but for Field S. The ALMA beam of $3\farcs1\times2\farcs1$  is shown in the bottom-left corner.}
  \label{fig2}
\end{figure*}

Figures \ref{fig1} and \ref{fig2} show the 3 mm continuum emission of G14.2 in each observed field without the primary beam correction.  The star signs and green crosses represent the positions of the YSOs and X-ray sources, respectively, identified by \citet{pov16}. The $\sim2\arcsec$ spatial resolution of the {\it Spitzer}/IRAC detector is similar to the ALMA synthesized beam, which facilitates the comparison between the location of cores and YSOs. The brightest regions in each field correspond to the hub-N and hub-S regions identified by \citet{bus13}. The ALMA images reveal several clumpy structures toward the hub regions.  
The peak fluxes of hub-N and hub-S are $\sim25$ and $\sim9$ mJy beam$^{-1}$, respectively. 
 The sensitivity (1$\sigma$) in each field is 0.2 mJy beam$^{-1}$, which corresponds to $\sim$ 0.28 $M_\odot$ and a column density of $N({\rm H_2})\sim2.4\times10^{22}$ cm$^{-2}$ ($\Sigma\sim0.1$ g cm$^{-2}$) at a dust temperature of 17 K \citep[kinetic temperature of hub-N and hub-S determined in ][]{bus13}. 
 Furthermore, assuming 0.1 pc width of filaments, the sensitivity of line mass corresponds to 52 $M_\odot$ pc$^{-1}$. \citet{bus13} estimated line mass of $74-328$ $M_\odot$ pc$^{-1}$ (average is 170 $M_\odot$ pc$^{-1}$ and $\Sigma\sim0.3$ g cm$^{-2}$) in the filaments. Therefore, we may be able to identify not only dense cores but also massive filaments.  However, we detect only compact dense cores. This may be explained by  the combined effect of resolving out extended emission from the filaments that have lengths longer than 1 pc ($\sim90\arcsec$) and insufficient sensitivity. In field N, \citet{bus13} identified two filaments (F10-E and F60-N) converging toward hub-N, while in the ALMA observations, we identify the highest column density peaks (cores) of both filaments. 
In field S, we mainly find the emission from the central region of the observed area corresponding to hub-S.

\subsection{Core properties}
To identify dense cores\footnote[1]{We will follow the nomenclature of \citet{zha09,zha14} and refer to clumps as an entity of $\lesssim1$ pc and a dense core as an entity of $0.01-0.1$ pc.
}, we apply the Clumpfind method \citep{wil94} to the ALMA 3 mm images with a threshold of 2$\sigma$ level  (2$\sigma$ interval steps are also used). 
This algorithm has been widely used for the identification for clumps and cores \citep[e.g.,][]{joh00,kir06,ike07,shi15}.
We identify a total of 48 cores (29 in the northern part and 19 in the southern part).
Figures \ref{fig1} and \ref{fig2} also show the positions (crosses and numbers) of the identified cores.

\citet{wil94} showed that the 2$\sigma$ noise level is the optimal threshold contour level to recover the clump and/or core structures. \citet{pin09} investigated the dependence of the threshold level from 3, 5, and 7$\sigma$ and stepsize from 2 to 20$\sigma$ on the core properties by testing both two- and three-dimensional clumpfind. They found that the power-law index of the CMF does not depend on threshold and stepsize compared to the fit uncertainties in the two-dimensional case, where emission is isolated. In our observations, the 3 mm continuum emission is relatively isolated and the clumpfind method is reliable to identify cores.    \citet{ike09} also showed that the power-law index remains unchanged for thresholds of $2-4$$\sigma$ and stepsizes of $2-3.5$$\sigma$ in the Orion A cloud. Therefore, we choose the threshold and stepsize of 2$\sigma$. Furthermore, we only consider cores having peak intensities above the $5\sigma$ noise level as robust detections.  

The spacing among the dense cores in these filaments is not constant and seems to be shorter when approaching the hub-N region.  
We also detect dense cores located in the filaments of F60-S and F60-C3 identified by \citet{bus13}. Recently, \citet{bus16} observed 1.3 mm continuum emission toward the hub-N and hub-S regions with the Submillimeter Array  (SMA). Their angular resolution is $\sim1\arcsec$ and they revealed some condensations toward these hub regions. We will compare the ALMA and SMA results in Sect. 3.2.

 We identify protostellar and prestellar cores by comparing the dense core positions with: i) the YSOs catalog of \citet{pov16}; ii) the presence or absence of 24 $\mu$m emission using the MIPSGAL survey \citep{car09}; and iii) presence of X-ray sources \citep[taken from][]{pov16}, regardless of IR detection. Assuming YSO ages of 0.1 Myr \citep{pov10} and a width of the velocity dispersion of 0.8 km s$^{-1}$, protostars may be able to move away from the core center as 0.4 km s$^{-1}$ $\times$ 0.1 Myr = 0.04 pc $\sim$ 4$\arcsec$. The width of the velocity dispersion of protostars is taken from N$_2$H$^+$ linewidth for prestellar cores candidates because it has been suggested that the stellar velocity dispersion is comparable to the gas motions  \citep{hac16}.  If these sources are inside a circle of radius 4$\arcsec$ from the center position of the core, the core is defined as protostellar. If no YSO is found within 4$\arcsec$ from the core center, we define the core as a prestellar candidate.
Following this definition, we identify 20 protostellar cores and 28 prestellar cores candidates.
 We should note that 0.1 Myr age is only applicable to the very youngest Stage 0/I YSOs but this is valid for most of the cores as they are mainly associated with Stage 0/I YSOs (as shown in Table \ref{alma}). We mention as a caveat that we cannot rule out the presence of deeply embedded low-mass protostars. They could be revealed in the future by deeper IR/X-ray observations or searching for molecular outflows.

The mass of the 3 mm dust cores, $M_{\rm core}$, are computed with
\begin{equation}
M_{\rm core}=\frac{F_\nu d^2}{\kappa B_\nu(T_d)}f_d,
\label{1}
\end{equation}
where $F_\nu$ is the observed flux in Jy, $d$ is the distance to the target, $B_\nu(T_d)$ is the Planck function at a dust temperature $T_d$, $\kappa$ is the dust opacity, and $f_d$ is the dust to gas mass ratio (assumed to be 100).  
We adopt the dust opacity
\begin{equation}
\kappa=0.9\Big(\frac{\lambda}{1.3 \rm\hspace{1mm} mm}\Big)^\beta \hspace{1cm} \rm cm^2 \hspace{1mm}g^{-1}.
\label{kappa}
\end{equation}
Using a dust mass opacity coefficient of 0.9 cm$^{2}$ g$^{-1}$ at 1.3 mm, which corresponds to
coagulated grains with thin ice mantles in cores with densities of $10^6$ cm$^{-3}$ \citep{oss94} and a dust emissivity index of $\beta=1.5$, we obtain $\kappa_{\rm 2.9 mm}=0.276$ cm$^2$ g$^{-1}$ \citep[e.g.,][]{ste15}.
For the dust temperature, we use the kinetic temperature, $T_{\rm kin}$, derived from NH$_3$ ($J,K$) $=$ ($1,1$) and $(2,2)$ emission lines \citep{bus13}. If the kinetic temperature is not available for  a core, we adopt the average value of 22 K for protostellar cores and 17 K for prestellar cores candidates in this region. As a result, the core mass ranges from 1.1 to 78 $M_\odot$. The typical size is $\sim3\farcs3$ (0.03 pc). The H$_2$ densities in these cores range from $10^5$ to 10$^7$ cm$^{-3}$.
Table \ref{alma} shows the coordinates, flux, mass, and temperature of the cores.  The associations with YSOs and/or X-ray sources are also reported in the table.  The ID numbers of YSOs and X-ray sources are adopted from \citet{pov16}.

\begin{deluxetable*}{p{9mm}cccccccccccc}
\tablewidth{0pt}
\tablecaption{Physical Parameters of dense cores \label{alma}}
\tablehead{
Sources &	R.A.	& Decl& $S_{\rm peak}$$\rm ^a$ & $S_{\rm int}$$\rm ^a$	& $T_{d}$$\rm ^b$	& Mass & Size$\rm ^c$	& r	& $\Delta v$$\rm ^d$	&YSO$\rm ^e$ & Xray$\rm ^f$ & $\alpha_{\rm virial}$ \\
& (h:m:s)	&(d:m:s) 	&(mJy beam$^{-1}$)	&(mJy)&(K) & ($M_\odot$) & ($\arcsec\times\arcsec$)&(pc)& (km s$^{-1}$)&  ID (Stage) & ID   &
}
\startdata
N-1 &18:18:12.31	&	-16:49:28.0	&	25.4	&	78.5	&		23	&	78	&	4.0	$\times$	5.1	&0.022	&$1.71\pm0.20$	&	 565(0/I)&		 &0.17\\
N-2 &18:18:11.50	&	-16:52:32.8	&	10.1	&	14.5	&		31	&	10	&	2.4	$\times$	2.8	&0.013	&$1.58\pm0.21$	&	548(0/I)&	 	&0.64\\
N-3 &18:18:13.06	&	-16:49:41.2	&	7.3	&	16.3	&		20	&	19	&	4.2	$\times$	3.7	&0.019	&$1.31\pm0.17$	&		&&	 0.36\\
N-4 &18:18:12.53	&	-16:49:22.0	&	5.1	&	12.3	&		22	&	13	&	4.7	$\times$	3.3	&0.019	&$1.42\pm0.19$	&	580(0/I)&	&	 0.62\\
N-5 &18:18:12.92	&	-16:49:29.6	&	3.6	&	19.8	&		19	&	24	&	4.2	$\times$	6.6	&0.025	&2 comp	&		&	 576	&$\dots$\\
N-6 &18:18:11.25	&	-16:52:37.2	&	3.5	&	5.5	&		22	&	5.8	&	3.1	$\times$	2.3	&0.013	&$1.35\pm0.19$	&		&&	 0.84\\
N-7 &18:18:12.67	&	-16:50:13.2	&	2.2	&	4.5	&		15	&	7.3	&	4.4	$\times$	5.0	&0.023	&$0.49\pm0.05$	&		&&	 0.16\\
N-8 &18:18:13.04	&	-16:48:17.6	&	4.7	&	8.0	&		17	&	11	&	2.0	$\times$	2.4	&0.010	&$0.74\pm0.16$	&		&&	 0.11\\
N-9 &18:18:12.40	&	-16:48:40.8	&	2.3	&	2.3	&	$\sim$17	&	3.2	&	$\dots$			&$\dots$	&$0.56\pm0.06$	&		&&	 $\dots$\\
N-10 &18:18:05.12	&	-16:53:42.4	&	6.6	&	7.4	&	$\sim$17	&	10	&	$\dots$			&$\dots$	&$\dots$	&		&&	 $\dots$\\
N-11 &18:18:12.53	&	-16:50:03.2	&	1.8	&	3.1	&		15	&	5.1	&	2.4	$\times$	2.7	&0.012	&$1.02\pm0.16$	&		&&	 0.54\\
N-12 &18:18:11.56	 &	-16:51:36.8	&	1.7	&	2.9	&	$\sim$22	&	3.0	&		$\dots$		&$\dots$	&$0.33\pm0.03$	&	544(0/I)	&&	 $\dots$\\
N-13 &18:18:06.85	&	-16:51:19.6	&	1.7	&	2.5	&	$\sim$22	&	2.6	&	1.5	$\times$	1.6	&0.007	&$0.55\pm0.07$	&	466(A)	&&	 0.18\\
N-14 &18:18:15.18	&	-16:49:20.8	&	1.7	&	4.5	&	$\sim$17	&	6.2	&	2.8	$\times$	5.0	&0.018	&$0.67\pm0.07$	&		&&	 0.28\\
N-15 &18:18:12.48	&	-16:49:58.8	&	1.4	&	2.4	&		16	&	3.7	&		$\dots$		&$\dots$	&$0.74\pm0.10$	&		&&	 $\dots$\\
N-16 &18:18:12.23	&	-16:49:36.0	&	1.4	&	6.0	&		22	&	6.3	&	5.9	$\times$	3.8	&0.023	&$1.02\pm0.15$	&		&&	 0.79\\
N-17 &18:18:05.65	&	-16:52:56.8	&	1.4	&	1.4	&	$\sim$22	&	1.5	&		$\dots$		&$\dots$	&$\dots$	&		&654&	 $\dots$\\
N-18	&18:18:08.05	&	-16:51:06.0	&	1.4	&	1.9	&	$\sim$22	&	2.0	&	0.7	$\times$	2.9	&0.007	&$\dots$	&	485(0/I)	&582&	 $\dots$\\
N-19	&18:18:12.76	&	-16:49:56.0	&	1.3	&	2.7	&		17	&	3.7	&	3.3	$\times$	2.3	&0.013	&$0.51\pm0.06$	&		&&	 0.19\\
N-20	&18:18:11.95	&	-16:49:33.6	&	1.3	&	3.4	&		21	&	3.8	&		$\dots$		&$\dots$	&$1.20\pm0.18$	&	554(II)	&491&	 $\dots$\\
N-21	&18:18:10.72	&	-16:49:49.2	&	1.2	&	3.4	&	$\sim$22	&	3.5	&	6.1	$\times$	3.4	&0.022	&$\dots$	&	529(II)	&512&	 $\dots$\\
N-22	&18:18:03.39	&	-16:52:29.6	&	1.2	&	1.5	&	$\sim$17	&	2.1	&	2.2	$\times$	1.9	&0.01	&$0.53\pm0.05$	&		&&	 0.27\\
N-23	&18:18:12.56	&	-16:49:47.6	&	1.1	&	3.3	&		19	&	4.1	&	2.6	$\times$	4.3	&0.016	&$1.26\pm0.17$	&		&&	 1.3\\
N-24	&18:18:13.73	&	-16:49:36.4	&	1.1	&	1.1	&		16	&	1.7	&		$\dots$		&$\dots$	&$0.59\pm0.07$	&		&&	 $\dots$\\
N-25	&18:18:14.71	&	-16:49:25.6	&	1.1	&	2.8	&	$\sim$22	&	2.9	&	2.2	$\times$	4.0	&0.014	&$0.56\pm0.08$	&		&420&	 0.32\\
N-26	&18:18:10.95	&	-16:49:44.0	&	1.1	&	1.1	&	$\sim$17	&	1.5	&		$\dots$		&$\dots$	&$\dots$	&		&&	 $\dots$\\
N-27	&18:18:12.23	&	-16:50:17.2	&	1.1	&	1.1	&	$\sim$22	&	1.1	&		$\dots$		&$\dots$	&$\dots$	&	564(0/I)	&487&	 $\dots$\\
N-28	&18:18:13.34	&	-16:49:24.0	&	1.1	&	1.1	&		16	&	1.6	&		$\dots$		&$\dots$	&$1.15\pm0.18$	&		& &$\dots$\\
N-29	&18:18:11.56	&	-16:49:26.4	&	1.0	&	1.2	&		20	&	1.4	&	1.3	$\times$	1.8	&0.008	&$0.58\pm0.06$	&	547(II)	&&	 $\dots$
\\\hline
S-1 &18:18:13.34	&	-16:57:23.8	&	8.7	&	21.7	&		21	&	27	&	4.0	$\times$	4.6	&0.020	&2 comp	&	585(0/I)	&&	 $\dots$\\
S-2&18:18:12.86	&	-16:57:20.2	&	6.8	&	18.6	&		22	&	22	&	4.1	$\times$	4.0	&0.019	&$1.73\pm0.21$	&		&&	 0.56\\
S-3&18:18:13.92	&	-16:57:11.4	&	5.7	&	16.1	&		17.5	&	24	&	3.4	$\times$	5.0	&0.020	&$0.96\pm0.14$	&		&449&	 0.15\\
S-4&18:18:12.42	&	-16:57:22.6	&	5.1	&	8.0	&		22	&	9.2	&	2.9	$\times$	2.3	&0.012	&$0.97\pm0.14$	&	562(II)	&&	 0.23\\
S-5&18:18:11.36	&	-16:57:28.2	&	3.0	&	4.1	&	$\sim$22	&	4.7	&	3.6	$\times$	2.6	&0.015	&$0.59\pm0.07$	&	539(0/I)	&&	 0.21\\
S-6&18:18:13.42	&	-16:57:11.0	&	2.3	&	9.2	&		15	&	17	&	3.0	$\times$	5.3	&0.019	&2 comp	&	589(0/I)	&&	 $\dots$\\
S-7&18:18:12.47	&	-16:57:26.6	&	2.2	&	5.3	&		19	&	7.2	&	3.7	$\times$	2.7	&0.015	&$1.16\pm0.16$	&		&&	 0.60\\
S-8&18:18:04.58	&	-16:58:47.4	&	2.0	&	2.0	&		14	&	3.7	&		$\dots$		&$\dots$	&$0.79\pm0.12$	&		&&	 $\dots$\\
S-9&18:18:09.16	&	-16:57:32.2	&	1.9	&	1.9	&	$\sim$17	&	2.7	&	$\dots$			&$\dots$	&$0.41\pm0.08$	&		&&	 $\dots$\\
S-10	&18:18:12.56	&	-16:57:15.8	&	1.6	&	3.2	&		20	&	4.1	&	2.9	$\times$	2.9	&0.014	&$1.07\pm0.19$	&		&&	 0.81\\
S-11 &18:18:13.92	&	-16:57:17.8	&	1.6	&	6.8	&		16	&	11	&	6.0	$\times$	3.5	&0.026	&2 comp	&		&445&	 $\dots$\\
S-12	&18:18:27.02	&	-16:55:27.0	&	1.6	&	2.3	&	$\sim$22	&	2.6	&	$\dots$	&$\dots$	&$0.58\pm0.06$	&	719(0/I)	&&	 $\dots$\\
S-13	&18:18:05.87	&	-16:58:53.0	&	1.5	&	3.7	&		16	&	6.2	&	1.5	$\times$	5.2	&0.014	&$0.66\pm0.08$	&		&&	 0.20\\
S-14	&18:18:15.93	&	-16:57:19.4	&	1.4	&	1.4	&		15	&	2.6	&		$\dots$		&$\dots$	&$0.64\pm0.08$	&		&&	 $\dots$\\
S-15	&18:18:05.39	&	-16:58:18.6	&	1.3	&	3.5	&		13	&	7.4	&	1.3	$\times$	5.4	&0.013	&$0.69\pm0.09$	&		&&	 0.17\\
S-16	&18:18:15.26	&	-16:56:30.2	&	1.3	&	1.6	&		12	&	3.9	&		$\dots$		&$\dots$	&$0.44\pm0.07$	&		&&	 $\dots$\\
S-17	&18:18:11.19	&	-16:57:18.2	&	1.2	&	3.6	&		17	&	5.6	&	5.7	$\times$	2.4	&0.018	&$1.14\pm0.19$	&		&&	 0.88\\
S-18	&18:18:13.03	&	-16:57:29.8	&	1.1	&	2.4	&		18	&	3.5	&	3.2	$\times$	3.2	&0.016	&$1.19\pm0.17$	&		&&	 1.3\\
S-19	&18:18:14.51	&	-16:57:14.2	&	1.0	&	2.7	&		15	&	4.8	&	6.0	$\times$	3.6	&0.022	&2 comp	&		&&	 $\dots$
\enddata
\tablenotetext{a}{Fluxes are corrected for the primary beam attenuation.}
\tablenotetext{b}{For dust temperature, we use the kinetic temperature from \citet{bus13}.  If temperature measurements are not available, we adopted the mean temperature for protostellar/prestellar cores.}
\tablenotetext{c}{Sizes are deconvolved with the beam size.}
\tablenotetext{d}{The linewidths are measured from N$_2$H$^+$ (J=1-0) ALMA observations (Chen et al. in prep).}
\tablenotetext{e}{YSO labels used in table 4 in \citet{pov16}.}
\tablenotetext{f}{X-ray sources used in table 6 in \citet{pov16}.}
\end{deluxetable*}

\subsection{Comparison with SMA observations}
\begin{figure*}[htbp]
  \begin{center}
  \includegraphics[width=16cm]{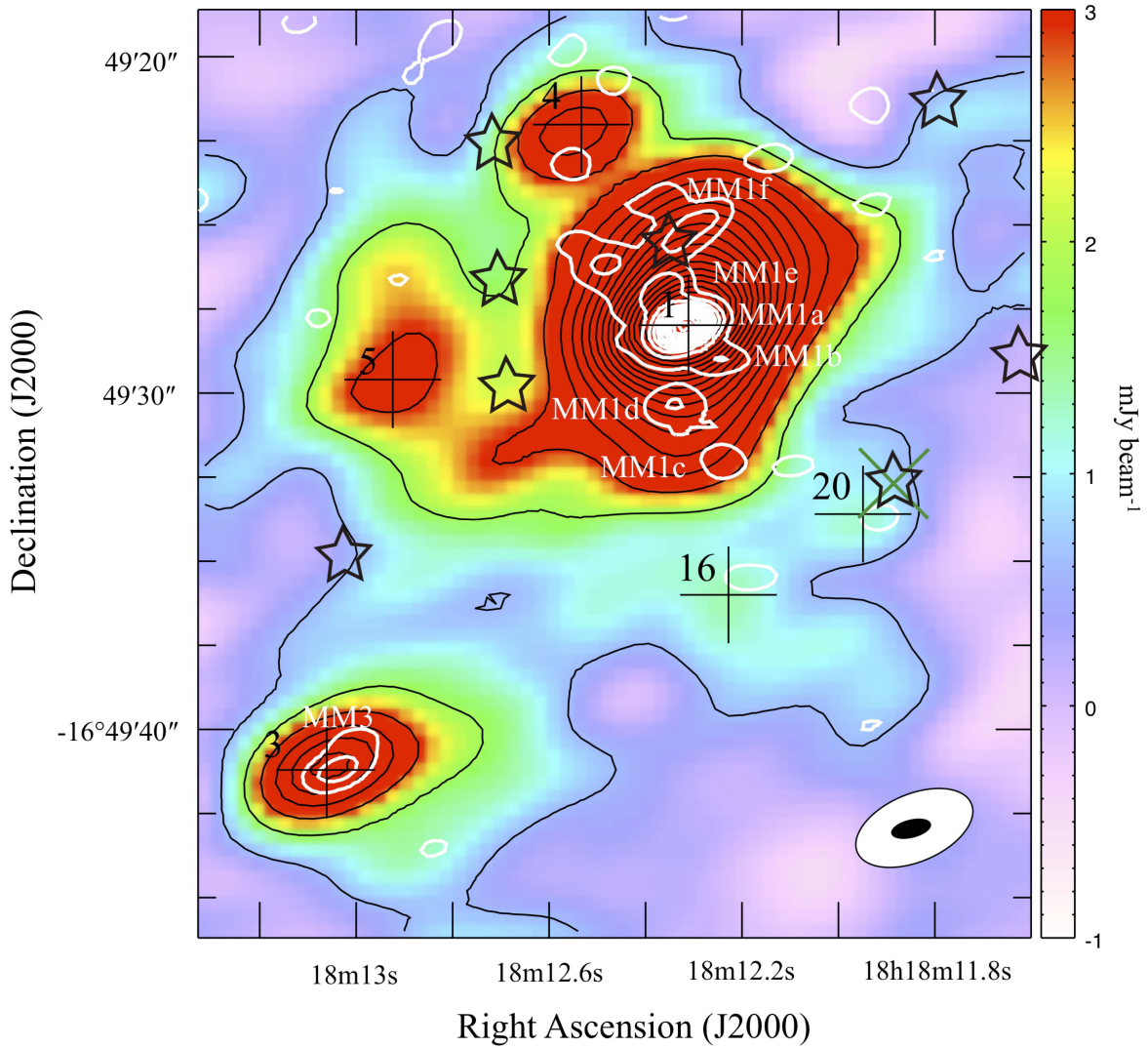}
  \end{center}
  \caption{Color image and black contour maps of the ALMA 3 mm continuum emission and  white contour maps of the SMA 1.3 mm continuum emission of G14.2 hub-N. The contours start at 3$\sigma$ and increase in steps of 6$\sigma$, where 1$\sigma$ is 0.2 mJy beam$^{-1}$ for ALMA and 1 mJy beam$^{-1}$ for SMA, respectively. Black crosses and numbers indicate the ALMA cores listed in Table \ref{alma}.  The 1.2 mm sources identified by \citet{bus16} are labeled. Star symbols and the green crosses represent the locations of the YSOs and the X-ray sources identified by \citet{pov16}, respectively. The ALMA beam of $3\farcs7\times2\farcs0$ and SMA beam of $1\farcs2\times0\farcs6$  is shown in the bottom-right corner.}
  \label{zoom_n}
\end{figure*}

\begin{figure*}[htbp]
  \begin{center}
  \includegraphics[width=16cm]{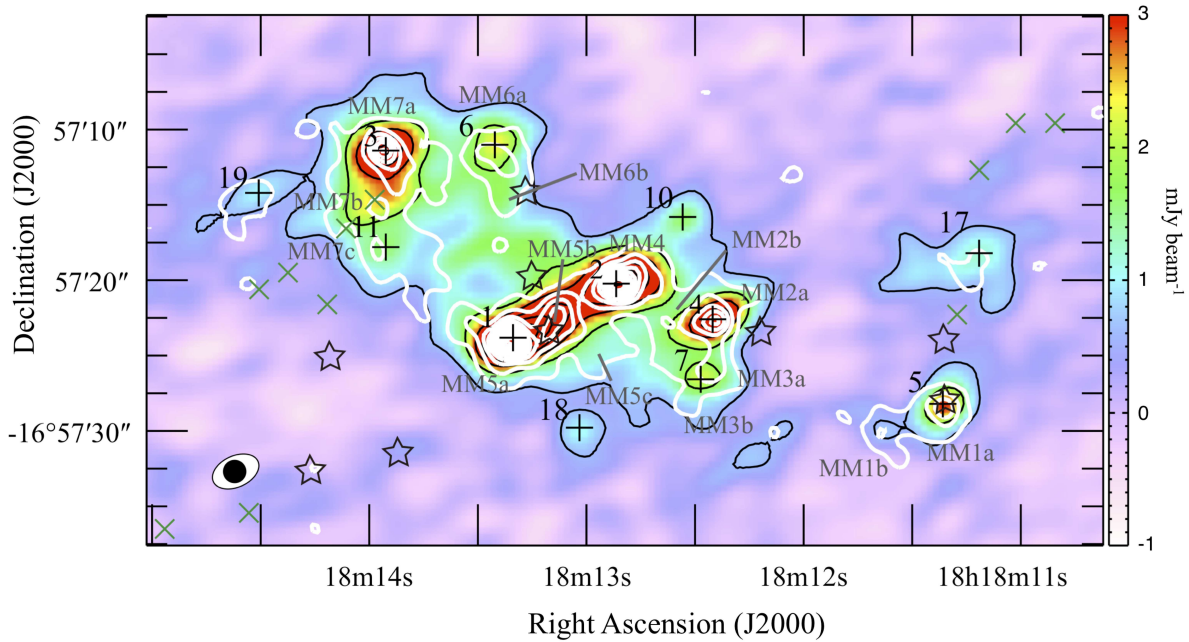}
  \end{center}
  \caption{Same as Figure \ref{zoom_n} but for hub-S.  The ALMA beam of $3\farcs1\times2\farcs1$ and SMA beam of $1\farcs5\times1\farcs4$  is shown in the bottom-left corner.}
  \label{zoom_s}
\end{figure*}

 Figures \ref{zoom_n} and \ref{zoom_s} show a zoom in of the ALMA images (color maps and black contours) and SMA 1.3 mm continuum images (white contours) toward hub-N and hub-S. 
Crosses and numbers mark the cores listed in Table \ref{alma}.
 We also labeled the SMA 1.3 mm sources identified by \citet{bus16}. 
  They observed 1.3 mm dust continuum emission toward both hub regions with SMA   at  $\sim$ 1$\arcsec$ angular resolution. The rms for the SMA observations of $\sim1$ mJy beam$^{-1}$ corresponds to 0.6 $M_\odot$ (6$\sigma$) at a dust temperature of 17 K. On the other hand, the rms for the ALMA observations of $\sim0.2$ mJy beam$^{-1}$ corresponds to 1.1 $M_\odot$ (4$\sigma$) at the same dust temperature. They revealed that both hubs fragment into several condensations, and therefore we can compare with our observations.

In Field-N, N-1 core is the most massive ALMA core with 78 $M_\odot$, but SMA observations revealed that N-1 consists of six condensations labelled as MM1a to MM1f  by \citet{bus16}. The brightest source MM1a is associated with a H$_2$O maser spot \citep{wan06}.  These six condensations have $\sim1\arcsec$ size and cannot be resolved in our ALMA observations. The N-3 prestellar core candidate coincides with SMA MM3, which shows no fragmentation in the SMA observations. N-4, 5, 20, and 29 protostellar cores and N-16 prestellar cores candidates were not detected by SMA. 

In the southern part, the ALMA and SMA images are similar except for the S-1 protostellar core. SMA observations identified three condensations (MM5a to MM5c) within the ALMA S-1 protostellar core. The brightest source MM5a is associated with a H$_2$O maser \citep{wan06}.  S-10, 17, 18, and 19 prestellar cores candidates were not detected by SMA, but note that S-17 and S-19 coincide with a 3 sigma contour level of the SMA image. In conclusion, the most massive protostellar cores of N-1 and S- 1 fragment into several condensations.  The finding of new cores undetected by SMA likely reflects the better dynamic range reached by ALMA. 

\subsection{Dynamical State of the Cores}
We investigate the dynamical state of the cores through virial analysis. By assuming a uniform density core, the virial mass is estimated as 
\begin{equation}
M_{\rm vir}=210 \times \left( \frac{r}{1 \; \mathrm{pc}} \right) \times \left( \frac{\Delta v}{1 \; \mathrm{km \, s^{-1}}} \right)^2 M_\odot,
\label{2}
\end{equation}
where $r$ is the FWHM radius and $\Delta v$ is the linewidth. The virial parameter, $\alpha_{\rm virial}$, is derived by $ \alpha_{\rm virial}=M_{\rm vir}/M_{\rm core}$.
We use ALMA observations of N$_2$H$^+$ $J=1-0$ to measure the linewidth toward the cores by performing hyperfine fitting to the N$_2$H$^+$ spectra. The 3 mm continuum and N$_2$H$^+$ emission were simultaneously observed and the velocity information will be reported in a forthcoming paper (Chen et al., in prep.). 
The average linewidths are  0.8 km s$^{-1}$ for prestellar and 1.0 km s$^{-1}$ for protostellar cores.
In general, all cores have virial parameters less than unity.   We find no clear differences in the virial parameters
between prestellar and protostellar cores (see table \ref{alma} for values of $\Delta v$, $r$, and virial parameter). 
The virial parameter is not calculated when: the cores are associated with multiple velocity components (4 out of 48 cores), the N$_2$H$^+$ emission is not detected (6 out of 48 cores), and/or cores are unresolved (15 out of 48 cores).

We also investigate the virial parameters at the scale of clumps and filaments (sizes $\sim$ 1pc for clumps and $\gtrsim1$ pc for filaments).  For this, we use NH$_3$ observations carried out with the VLA and Effelsberg 100 m telescope and 870 $\mu$m continuum observations carried out with  APEX \citep{bus13}. The spatial resolution of the NH$_3$ and 870 $\mu$m emission is $\sim8\arcsec$ and 22$\arcsec$ equivalent to 0.08 and 0.21 pc, respectively. We apply the Clumpfind algorithm to the 870 $\mu$m images with a threshold of $2\sigma$ and an interval of $2\sigma$. We only consider clumps having intensities higher than 0.4 Jy beam$^{-1}$, corresponding to 5$\sigma$ above the noise level for the 870  $\mu$m images. 
We consider in the analysis only the clumps located within the ALMA fields. The clump mass is derived by using equations (\ref{1}) and (\ref{kappa}). The dust opacity $\kappa$ at 870 $\mu$m is derived to be 1.64 g cm$^{-2}$ by using equation (\ref{kappa}). The properties of the clumps are listed in Table \ref{apex}. We use the NH$_3$ ($J,K$) $=$ $(1,1)$ line to measure the linewidth toward the clumps.
 \citet{mac88} suggested that the coefficient of the virial mass depends on the density profile, $\rho\propto r^a$, with the power-law index $a$. For uniform density ($a=0$) the coefficient is 210, while for a power-law index ($a=-2$) the coefficient is 126  (see equation \ref{2}). Therefore, our virial parameters for cores and clumps correspond to upper limits.

We also refer to the virial parameters of the parent filaments derived by \citet{bus13}. They investigated the stability of the filaments by estimating the virial parameter, which is slightly different from the core stability in terms of the coefficient \citep[210 for cores and 332 for filaments;][]{ber92}.
In filaments, the virial mass is estimated as
\begin{equation}
M_{\rm vir}=\frac{2\sigma^2}{G}\ell\sim332 \times \left( \frac{\ell}{1 \; \mathrm{pc}} \right) \times \left( \frac{\Delta v}{1 \; \mathrm{km \, s^{-1}}} \right)^2 M_\odot,
\label{3}
\end{equation}
where $\ell$ is the length of the filaments.

\begin{deluxetable*}{ lllccccccccc }
\tabletypesize{\scriptsize}
\tablecaption{Physical Parameters of dense clumps \label{apex}}
\tablewidth{0pt}
\tablehead{
Clump	&	R.A.	&	Decl	&	$S_{\rm peak}$		&	$S_{\rm int}$	&	$T_d$$\rm ^a$	& Mass & Size	& r	&	$\Delta v$$\rm ^b$&	$\alpha_{\rm virial}$	& Field	 \\
	&	(h:m:s)	&	(d:m:s) 	&	(Jy beam$^{-1}$)	&	(Jy)	&	(K)	 & ($M_\odot$) & ($\arcsec\times\arcsec$)	&(pc)	&(km s$^{-1}$)&		&
}
\startdata
1	&	18:18:12.6	&	-16:49:34	&	6.41	&	35.6	&	17	&	1400	&	53	$\times$	85&0.32	&$2.57\pm0.29$	&0.33&hub-N	\\
2	&	18:18:13.2	&	-16:57:22	&	4.47	&	25.3	&	17	&	960	&	79	$\times$	48&0.30	&$3.83\pm0.99	$&0.95&hub-S	\\
3	&	18:18:11.2	&	-16:52:34	&	1.48	&	5.1	&	18	&	180	&	39	$\times$	37&0.18	&$1.59\pm0.20	$	&0.55&N	\\
4	&	18:18:07.9	&	-16:51:22	&	1.23	&	8.1	&	13	&	450	&	39	$\times$	55&0.22	&$1.73\pm0.30	$	&0.31&N	\\
5	&	18:18:05.7	&	-16:58:18	&	1.11	&	7.7	&	11	&	620	&	50	$\times$	53&0.25	&$1.88\pm0.28$		&0.30&S	\\
6	&	18:18:05.9	&	-16:58:54	&	1.01	&	4.6	&	11	&	370	&	51	$\times$	29&0.18	&$1.81\pm0.25$		&0.35&S	\\
7	&	18:18:11.5	&	-16:51:42	&	1.01	&	5.2	&	18	&	180	&	52	$\times$	51&0.25	&$2.54\pm0.27$		&1.9&N	\\
8	&	18:18:04.8	&	-16:51:42	&	0.95	&	6.3	&	13	&	360	&	35	$\times$	61&0.22	&$1.94\pm0.23$		&0.50&N	\\
9	&	18:18:19.6	&	-16:55:58	&	0.58	&	3.9	&	11	&	310	&	50	$\times$	54&0.25	&$2.79\pm0.32$		&1.3&S	\\
10	&	18:18:05.1	&	-16:53:46	&	0.57	&	1.6	&	18	&	55	&	28	$\times$	45&0.17	&$1.70\pm0.21$		&1.9&N	\\
11	&	18:18:21.0	&	-16:56:38	&	0.49	&	2.5	&	11	&	200	&	39	$\times$	40&0.19	&$2.99\pm0.28$		&1.8&S	\\
12	&	18:18:25.2	&	-16:56:30	&	0.49	&	1.6	&	11	&	130	&	24	$\times$	31&0.13	&$2.39\pm0.48$		&1.2&S	\\
13	&	18:18:08.7	&	-16:56:30	&	0.49	&	2.5	&	11	&	200	&	59	$\times$	32&0.21	&$2.03\pm0.23$		&0.90&S	\\
14	&	18:18:25.7	&	-16:54:02	&	0.44	&	1.0	&	11	&	78	&	29	$\times$	23&0.12	&$1.87\pm0.21$		&1.2&S	\\
15	&	18:18:27.1	&	-16:55:30	&	0.43	&	1.4	&	11	&	110	&	37	$\times$	25&0.15	&$1.72\pm0.20$		&0.82&S	\\
16	&	18:18:04.5	&	-16:59:50	&	0.4	&	2.4	&	11	&	190	&	35	$\times$	62&0.22	&$1.14\pm0.18$		&0.32&S	\\
17	&	18:18:09.8	&	-16:53:58	&	0.39	&	1.3	&	18	&	45	&	28	$\times$	32&0.14	&$1.54\pm0.19$		&1.6&	N\\
18	&	18:18:01.5	&	-16:57:42	&	0.33	&	0.9	&	11	&	68	&	34	$\times$	27&0.14	&$2.26\pm0.33$		&2.3&S	
\enddata
\tablenotetext{a}{For dust temperature, we use the kinetic temperature from \citet{bus13}. }
\tablenotetext{b}{The linewidths are measured from NH$_3$ ($J,K=1,1$)  observations \citep{bus13}.}
\end{deluxetable*}

\begin{figure}[htbp]
  \begin{center}
  \includegraphics[width=8cm]{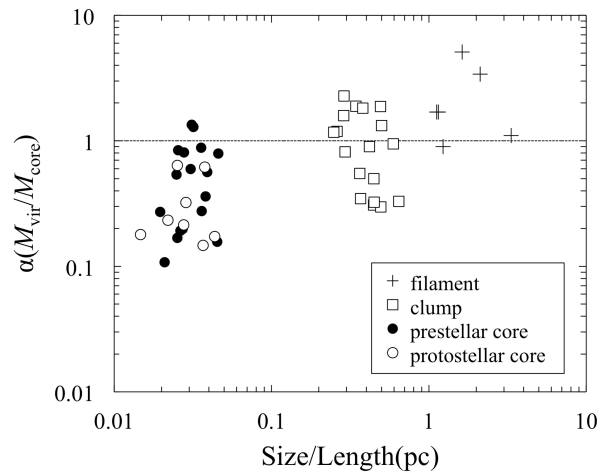}
  \end{center}
  \caption{Virial parameter, $\alpha$, versus sizes/lengths. Filled and open circles indicate the prestellar and protostellar cores identified with ALMA. Open squares indicate the APEX clumps. Crosses indicate the parent filaments.}
  \label{virial}
\end{figure}

Figure \ref{virial} shows the virial parameters against the clump/core sizes and filament lengths. We find a tendency showing that $\alpha_{\rm virial}$ decreases with decreasing size scales.  Almost all ALMA cores have $\alpha_{\rm virial}$ of less than 1 and the APEX clumps have $\alpha_{\rm virial}$ of $0.3-2.3$ (the average value is $\alpha_{\rm virial}=1.0$). Their parent filaments also show $\alpha_{\rm virial}=0.9-5.1$  (the average value is $\alpha_{\rm virial}=2.3$).  
The presence of external pressure may further reduce the virial parameters. Figure \ref{virial} suggests that APEX clumps and their parent filaments are in virial equilibrium on average. However, ALMA cores show virial parameters of less than 1, suggesting that the dense cores are not in virial equilibrium unless magnetic field support is important. 
We should note that the most massive clumps, associated with hub regions, also show virial parameters of less than 1. Therefore, these clumps may be able to supply material to the cores. Cores are likely located in privileged positions where the gravitational potential can significantly change the mass of the cores.

In numerical studies, there
are mainly two scenarios for the formation of high-mass stars. One is the turbulent core accretion model, in which
massive cores are supported by turbulence and collapse under gravitational pull
\citep[e.g.,][]{mck03}. The other model suggests that molecular cloud cores are
dynamically formed from large-scale flows \citep{mac04}. 
Some cores accrete mass faster than others due to their privileged position in the global gravitational potential, without well-defined massive bound structures of high-mass prestellar cores, as suggested by the turbulent core accretion model. This scenario is called the competitive accretion model \citep{bon97,bon01}. 
It is suggested that high-mass star-forming regions often show low values of $\alpha_{\rm virial}$ \citep{pil11,li13,tana13,san13,fos14,lu14,zha15}, which may follow ``competitive accretion'' \citep{kau13}.
Our observations may also support this scenario.

The virial analysis performed above neglects effects of magnetic fields, which can increase the virial mass. 
If magnetic fields are included in the virial
equation, the following expression holds
\begin{equation}
M_{\rm B,vir}=3\frac{R}{G}\left( \frac{5-2a}{3-a} \right)\left(\sigma^2+\frac{1}{6}\sigma_A^2 \right),
\label{mag}
\end{equation}
where $\sigma$ is the velocity dispersion ($\Delta v=\sqrt{8\ln2}\hspace{2mm}\sigma$) and $\sigma_A$ is the the Alfven velocity which can be determined from
\begin{equation}
\sigma_A=\frac{B}{\sqrt{4\pi\rho}},
\label{alf}
\end{equation}
where $B$ is the magnitude of the magnetic field and $\rho$ is the mass density.
To make the virial parameters close to 1, magnetic fields of $1-20$ mG (average value of 4 mG) would be needed.
In \citet{bus16}, they reported the magnetic fields are $\sim1$ mG in hub-N and 1.5 mG in hub-S. These values were obtained from near-infrared polarization measurements \citep{san16}, which traces more diffuse gas around the hub-filament systems but were extrapolated to the clump densities. 
\citet{cru10} suggest the maximum strength for magnetic field as $B_{\rm max}\propto n^{0.65}$ for $n>300$ cm$^{-3}$.
Following this relation, the most probable maximum strength for the magnetic field is $\sim1$ mG at a density of 10$^6$ cm$^{-3}$, which is consistent with the reported values.
In recent observations of dust polarization in a large sample of massive star-forming regions, \citet{zha14} found that magnetic fields play an important role during the collapse of massive clumps and the formation of massive cores.  The survey also reported the magnitude of the magnetic field of order of $1-10$ mG \citep{gir09,gir13,tang09,qiu13,qiu14}.
However, all estimates of magnetic field strengths have been made for star-forming cores and clumps. No measurements have been done for prestellar regions.
Even if we assume a magnetic field of 1 mG in hub-N and 1.5 mG in hub-S, the virial parameters for prestellar cores candidates are still smaller than unity in the hub regions.

\subsection{Caveats of Core Sizes and  Extended Emission}
ALMA observations of cores in IRDCs \citep[e.g.,][]{per13,zha15,tan16} and high-mass star-forming regions \citep[e.g.,][]{sak13,guz14,hig15} show cores diameters smaller than 
0.06 pc, which is similar to the sizes obtained in our work. 
Using the clumpfind method, we identified ALMA compact cores whose sizes (FWHM) are $\sim0.03$ pc ($\sim3\farcs3$).
Remarkably, the work of \citet{guz14} toward the high-mass young stellar object, G345.4938+0.14677, with a sensitivity 10 times better than our observations only finds cores smaller than 2$\farcs$5 ($\sim0.02$ pc), even though the shortest baselines were 21 m corresponding to LAS of $\sim13\arcsec$ ($\sim0.11$ pc) at 3 mm, following equation A5 of \citet{pal10}.
Since the LAS of \citet{guz14} ($\sim13\arcsec$) is much larger than the size of the cores ($\sim2\farcs5$), the cores seem to be intrinsically compact and ALMA should not be missing a significant amount of extended emission.
Therefore, we assume that the typical dense core size is $\lesssim0.06$ pc and our ALMA observations recover most of the flux from the dense cores in G14.2.  This is  the assumption for the rest of the paper. 
If the real size of the cores is larger than our estimated size, our final conclusions should be taken with caution.

\section{Discussion}
\subsection{Core Mass Function}
Following the IMF, the CMF is usually expressed 
\begin{equation}
\frac{dN}{dM}\propto M^{-\gamma},
\label{diff}
\end{equation}

and in the cumulative form of
\begin{equation}
N(>M) = k M^{-\alpha}+c,
\label{cum}
\end{equation}
where $\alpha=\gamma-1$. The differential expression of equation (\ref{diff}) allows us to find the turnover of the slope.
However, the shape is sensitive to the mass bins adopted.
On the other hand, the cumulative form of equation (\ref{cum}) does not strongly depend on the mass bins and allow us to fit the power-law, $\alpha$.
To construct the  CMF, we only use the 28 prestellar core candidates.

Figure \ref{cmf} shows the cumulative mass function, $N(>M)$, $N$ is the number of sources with mass larger than $M$.  The dashed lines represent the slope of the Salpeter IMF ($\alpha=1.35$). As shown in Figure \ref{cmf}, we can see that the cumulative CMF is similar to the Salpeter IMF. The power-law index $\alpha$ of the CMF is derived to be $1.6\pm0.7$ using the mass range from 2.4 to 14 $M_\odot$, taking into account the counting uncertainties (poisson noise) of $N^{1/2}$.  We found that the CMF is similar to the IMF, but it has a deficit of cores at the high-mass end. The lowest mass bin of 1.5 $M_\odot$ is affected by the detection limit.
 We have found that S-7 prestellar core candidate is resolved into compact condensations with SMA. Some dense cores may further fragment and smaller condensations might exist. If this is the case, the CMF could become steeper.

\subsection{Star formation in G14.2}
 Assuming that the final stellar mass function in G14.2 follows the Salpeter IMF, \citet{pov16} suggest that the massive cores are still in the process of accreting sufficient mass to form massive clusters hosting O stars and  high-mass stars may be formed later (as also has been speculated in IRDC G34.43+00.24: \citealt{fos14}).

On the contrary, G28.34+0.06 shows a significant deficit of low-mass core population around  high-mass protostellar cores \citep{zha15}.  
This disagreement can coincide if the low-mass stars at early times are formed in the outskirts of clumps, although this scenario was already suggested as less likely in G28.34+0.06 \citep{zha15}. Using deep near-infrared observations with adaptive optics, \citet{fos14} discovered a  distributed population of low-mass protostars within the filamentary IRDC G34.43+00.24 (located between clumps). It could also be the case that deep near-infrared observations could reveal low-mass protostars in the outskirts of G28.34+0.06, which could follow global collapse and move closer to the center of the clump.

To estimate the  ``expected'' IMF and maximum stellar mass in the G14.2 region observed with ALMA, we estimate the total gas mass and use the empirical relation of \citet{lar03}. 
Using the 870 $\mu$m continuum images obtained by APEX (Figure \ref{map}), the total mass of G14.2 within the ALMA fields is $\sim$5,800 $M_\odot$, assuming a temperature of 13 K (the average temperature of the APEX clumps) and a dust opacity of 1.64 g cm$^{-1}$ at 870 $\mu$m by using equation (\ref{kappa}). 

\begin{figure}[htbp]
  \begin{center}
  \includegraphics[width=8cm]{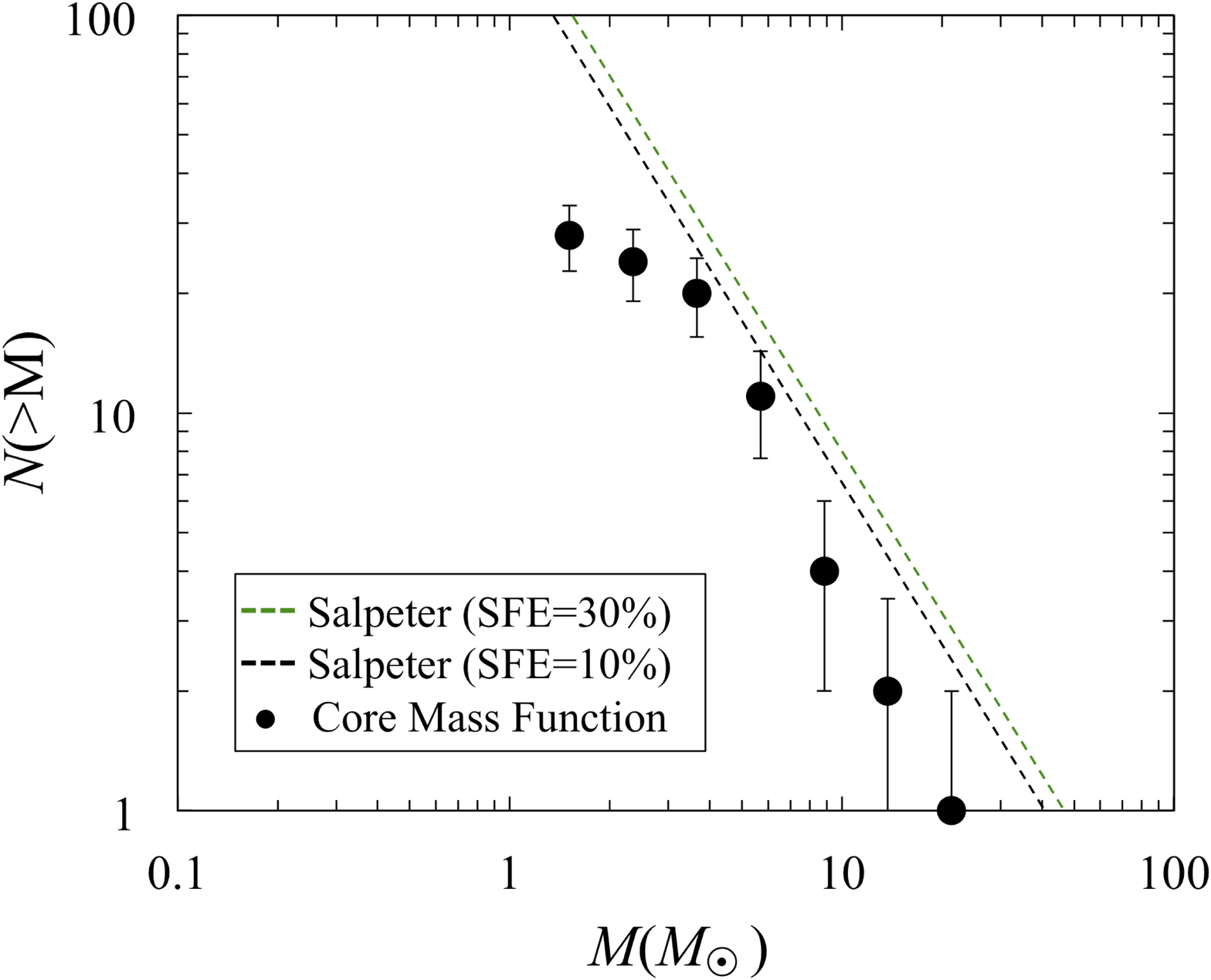}
  \end{center}
  \caption{Cumulative mass function of prestellar core candidates identified in G14.2 with ALMA. The data are plotted in logarithmic scales. The black and green dash lines represent the expected cumulative Salpeter IMF assuming the SFE of 10\% and 30\%, respectively.}
  \label{cmf}
\end{figure}

 If 10\% or 30\% of the total mass will end up as stellar mass (cluster star formation efficiency, SFE, of 10\% and 30\%), the maximum stellar mass is estimated to be $\sim21$ and 34 $M_\odot$, using the empirical relation $M_{\rm max}=1.2 M^{0.45}_{\rm cluster}$ where $M_{\rm max}$ is the maximum stellar mass and $M_{\rm cluster}$ is the cluster's mass \citep{lar03}. Because we have information of the current protostellar content in G14.2, we can make a more accurate estimation of the maximum stellar mass.    In \citet{pov16}, there are 334 YSOs in the {\it Chandra} observing field of view and 139 YSOs are located inside the ALMA fields. Thus, 41.6\% (139/334) of the YSOs are found within the ALMA fields. The current total stellar mass of highly obscured and clustered YSOs with $A_V>$20 mag is estimated  to be  4,300 $M_\odot$  \citep{pov16}. We estimate the total stellar mass associated with dense gas within the ALMA fields as $4,300M_\odot\times0.416=1,789$ $M_\odot$. Then, we add the current stellar mass, 1,789 $M_\odot$, and the expected final stellar cluster mass, SFE$\times$5,800 $M_\odot$.
 Taking these values, we derive the maximum stellar mass to be {\bf 40 and 47 $M_\odot$, respectively}. We note that these values are 1.8 and 2.1 times larger than the most massive prestellar core.
From the maximum stellar mass, we expect a Salpeter IMF in this region as shown in Figure \ref{cmf} in the dashed lines.
In the cumulative IMF, the maximum stellar mass starts at the 40 and 47 $M_\odot$  (for $N=1$)  and have a  power-law index of $\alpha=1.35$. 
From this figure, the prestellar cores can recover the deficit  at the high-mass end if,  in addition to completely convert their whole mass into stars, they continue growing.

Another possibility to recover the deficit of YMF at the high-mass end is that YSOs are grown by accreting the surrounding material, shifting the current YMF more top-loaded to higher masses. This is suggested by theoretical models of global hierarchical collapse of molecular clouds \citep{zam12}.
Indeed, some YSOs are embedded in protostellar cores with  a large mass reservoir and can grow by accretion.
 \citet{bus16} identified 26 compact sources in Hub-N and Hub-S, ranging in mass from $<1$ to 18 $M_\odot$ (they resolved in smaller fragments the two more massive cores detected by ALMA). None of these sources is currently massive enough to form an O-type star.
If this is the case, the YSOs would need to efficiently accrete more of the remaining gas  to recover the Salpeter IMF from outside of cores, which is also suggested by \citet{zha11},\cite{wan11}, and \citet{pil11}.

Whether the Salpeter IMF in G14.2 is recovered by the formation of high-mass stars from the current prestellar cores or accretion from the current YSOs (or both), a high star-formation efficiency at core scales is needed.
Recent observations \citep[e.g.,][]{alv07,pal15} and simulations \citep[e.g.,][]{mat00} suggest SFE $\lesssim$ 50\%. 
Considering that molecular outflows reduce star-formation efficiencies, in order to obtain values close to 100\%, a continuous gas replenishment to the core from the surrounding clump is necessary.

Our observations cannot firmly support or refute current theories of high-mass star formation.  However, given the extremely high SFEs necessary to produce high-mass stars from the prestellar and/or protostellar cores in G14.2, the star-formation process in this region may be more consistent with a global hierarchical collapse of the cloud, where competitive accretion takes place \citep[e.g.,][]{bon04,wan10}. In this scenario, low-mass stars form first simply because the density fluctuations of the cloud which first collapse are those of smaller scale and mass \citep[e.g.,][]{zam12}. Simultaneously with the formation of low-mass stars, gas and stars feel the potential well of the entire cloud and flow towards the center where the potential well is deepest. High-mass stars form later from low-mass stars, which in turn gain mass from the surrounding cloud. Thus, the full IMF will not adopt the Salpeter's form until the process of accretion inside a collapsing cloud has already finished. 
 
 We should note that the two most massive prestellar cores (N-3 and S-2) have masses of 19 and 22 $M_\odot$, respectively.
However, these cores still need to increase their masses to recover the Salpeter IMF even though high-mass stars may be formed in these massive prestellar cores.
The question of the existence of threshold conditions for the mass surface density ($\Sigma$) to form high-mass stars is still unclear, but some observations suggest a threshold of $\Sigma\sim0.3$ g cm$^{-2}$ \citep{lop10,but12,tan14}.
Our ALMA sensitivity is sufficient to observe such high-mass star-forming regions above 0.3 g cm$^{-2}$ (consistent with the average surface density of the G14.2 filaments) in spite of the fact that the massive cores are very small (size $\sim0.04$ pc).

Therefore, we suggest the following star-formation scenario in G14.2. Prestellar cores have a mass distribution similar to the Salpeter IMF. 
Low- and intermediate-mass YSOs are formed, resulting in a YMF steeper than the Salpeter IMF (and CMF). 
Later, assuming a high SFE, high-mass stars are formed either from collapse of massive prestellar cores which accrete a significant amount of surrounding gas or prolonged accretion from protostellar cores to the current YSOs (or both), resulting in the Salpeter IMF.

\section{conclusions}
We present 3 mm continuum observations carried out with ALMA toward the IRDC G14.225-0.506.
This survey covers the two hub-filament systems.
The main findings of this work are summarized as follows;\\

1. We identified 48 dense cores in this region with the clumpfind method.
The mass ranges from  1.1 to 78 $M_\odot$.
 The two most massive cores are protostellar and fragment in smaller condensations at higher angular resolution with SMA \citep{bus16}. We found that 20 dense cores are associated with protostellar activity as revealed by the by {\it Spitzer Space Telescope} or {\it Chandra X-ray Observatory} and 28 dense cores are not associated with known IR or X-ray emission.
Furthermore, using 870 $\mu$m continuum emission obtained with APEX, we also identified 18 clumps hosting these cores.\\

2. Using ALMA N$_2$H$^+$ and VLA/Effelsberg NH$_3$ molecular lines, we analyzed the virial equilibrium of cores and clumps.
As a result, we found a trend that the virial parameter decreases with decreasing scales from filaments to clumps and to cores.
The virial parameters of the dense cores are between 0.1 and 1.3, indicating that the internal support is insufficient and cores are undergoing dynamical collapse.
The clumps located in the hubs or massive clumps also present virial parameters of less than 1. 
Therefore, clumps may be able to supply material to the cores located in privileged positions of the global gravitational potential.\\

3.  The cumulative Core Mass Function for the prestellar cores had a power law index ($\alpha=1.6$), suggesting that the CMF has similar shape to the Salpeter IMF.  The prestellar core masses range from 1.5 to 22 $M_\odot$.  Previous results suggest that massive O-tpye stars has not be produced yet in this region. We suggested that high-mass stars can be formed in the prestellar cores by accreting a significant amount of the surrounding gas and/or  low-mass protostars  grow in the cores, fed by the filaments.

\acknowledgments

We thank Matthew S. Povich for his  valuable constructive comments that have improved the presentation of the paper.
S.O. is   financially   supported   by   a   Research   Fellowship   from the  Japan  Society  for  the  Promotion  of  Science  for  Young Scientists.
G.B. acknowledges the support of the Spanish Ministerio de Economia y Competitividad (MINECO) under the grant FPDI- 2013-18204. G.B. is also supported by the Spanish MINECO grant AYA2014- 57369-C3-1-P.
A.P. appreciates insightful discussions with Enrique V{\'a}zquez-Semadeni and Javier Ballesteros-Paredes. A.P. acknowledges financial support from UNAM-DGAPA-PAPIIT IA102815 grant, M\'exico.
This paper makes use of the ALMA data: ADS/JAO.ALMA 2013.1.00312.S. 
ALMA is a partnership of ESO (representing its member states), NSF (USA) and NINS (Japan), together with NRC (Canada), NSC and ASIAA (Taiwan),   and  KASI  (Republic  of  Korea),   in  cooperation with the  Republic  of Chile.   The joint  ALMA Observatory  is operated  by  ESO,  AUI/NRAO,  and  NAOJ. 
Data analysis were carried out on common use data analysis computer system at the Astronomy Data Center, ADC, of the National Astronomical Observatory of Japan.



{\it Facilities:} \facility{No:ALMA}.

\end{document}